\documentclass[prl,twocolumn,showpacs,floats,floatfix]{revtex4}
\usepackage{amsmath,amssymb,bm,graphicx}

\begin{document}

\title{Nonlinear screening and ballistic transport in
a graphene $p$-$n$ junction}

\author{L. M. Zhang and M. M. Fogler}


\affiliation{Department of Physics, University of California San
Diego, La Jolla, 9500 Gilman Drive, California 92093}

\date{\today}

\begin{abstract}

We study the charge density distribution, the electric field profile,
and the resistance of an electrostatically created lateral $p$-$n$
junction in graphene. We show that the electric field at the interface
of the electron and hole regions is strongly enhanced due to limited
screening capacity of Dirac quasiparticles. Accordingly, the junction
resistance is lower than estimated in previous literature.

\end{abstract}


\maketitle

\begin{figure}[t]
\centerline{
\includegraphics[width=2.9in]{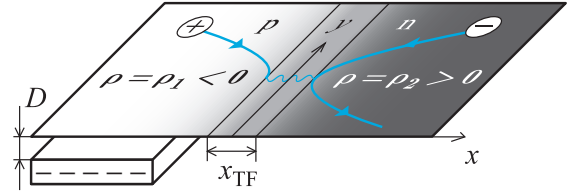}
}
\caption{\label{Fig:Gate} Device geometry. The semi-infinite gate on the
left side (beneath the graphene sheet) controls the density drop $\rho_2
- \rho_1$ across the junction, while another infinite ``backgate'' above
the sheet (not shown) fixes the density $\rho_2$ at far right. The
smooth curves with the arrows depict typical ballistic trajectories of an
electron ($-$) and a hole ($+$). The wavy curve symbolizes
their recombination via quantum tunneling.
}
\end{figure}

Unusual electron properties of graphene are an active topic of
fundamental research and a promising source of new
technology~\cite{Novoselov_04}. A monolayer graphene is a gapless
two-dimensional (2D) semiconductor whose quasiparticles (electrons and
holes) move with a constant speed of $v \approx 10^6\, {\rm m}/{\rm s}$.
The densities of these ``Dirac'' quasiparticles can be controlled by
external electric fields. Recently, using miniature gates, graphene
$p$-$n$ junctions (GPNJ) have been realized
experimentally~\cite{Huard_07}. In such electrostatically created GPNJ
the electron density $\rho(x)$ changes \emph{gradually\/} between two
limiting values, $\rho_1 < 0$ and $\rho_2 > 0$, as a function of
position $x$. This change occurs over a lengthscale $D$ determined by
the device geometry. For a junction created near an edge of a wide gate
(Fig.~\ref{Fig:Gate}), $D$ is of the order of the distance to this gate.

In addition to opening the door for novel device applications,
the study of transport through GPNJ may also test intriguing
theoretical ideas of Klein tunneling~\cite{Katsnelson_06} and
Veselago lensing~\cite{Cheianov_07}. Klein tunneling (also known as
Landau-Zener tunneling in solid-state physics) in graphene is both
similar and different from its counterpart for massive Dirac
quasiparticles, which was studied much earlier in semiconductor
tunneling diodes. In such a diode there is a uniform electric field
$F_{pn}$ in the gapped region and the quasiparticle tunneling
probability is given by $T = \exp(-\pi \Delta^2 /\, e \hbar v
|F_{pn}|)$~\cite{Kane_69}. The \emph{single-particle\/} problem for a
massless case is mathematically equivalent, except the role of the gap
$\Delta$ is played by $\hbar v k_y$. Integrating $T(k_y)$ over the
transverse momentum $k_y$ to get conductance and then inverting it,
one finds the ballistic resistance $R$ per unit width of the GPNJ to
be~\cite{Cheianov_06}
\begin{equation}
\label{eq:R_from_F}
R = (\pi / 2) (h / e^2) \sqrt{\hbar v \,/\, e |F_{pn}|}\,.
\end{equation}
Therefore, the absence of a finite energy gap prevents $R$ from becoming
exponentially large. This makes GPNJs orders of magnitude less resistive
than tunneling diodes, in a qualitative agreement with
experiment~\cite{Huard_07}. However as soon as one starts to think about
quantitative accuracy, one quickly realizes that many-body effects make
Eq.~(\ref{eq:R_from_F}) dubious. In particular, the model of a uniform
electric field, which is crucial for the validity of
Eq.~(\ref{eq:R_from_F}), is simply not correct in real graphene devices.
First, the absence of a gap and second, the nonlinear nature of
screening due to strong density gradients in a GPNJ invalidate such an
approximation. Furthermore, since the massless electrons and holes can
approach the $p$-$n$ interface very closely, their coherent
recombination takes place inside a ``Dirac'' strip of some
characteristic width $x_\text{TF}$ (Fig.~\ref{Fig:Gate}) whose
properties inherit the properties of the Dirac vacuum. Those can be
profoundly affected by strong Coulomb interactions and presently remain
not fully understood. This suggests that the problem of transport across a
GPNJ is still very much open.

Below we show that a controlled analysis of this problem is nevertheless
possible if one treats the dimensionless strength of Coulomb
interactions $\alpha = {e^2}/{\kappa_0 \hbar v}$ as a small parameter.
Here $\kappa_0$ is the effective dielectric constant. Small $\alpha$ can
be realized using HfO$_2$ and similar large-$\kappa_0$ substrates or
simply liquid water, $\kappa_0 \sim 80$. For graphene on a conventional
SiO$_2$ substrate, $\kappa_0\approx 2.4$ and $\alpha \approx 0.9$. For
such $\alpha$ we expect nonnegligible corrections to our analytic
theory, perhaps, $25\%$ or so.

Our main results are as follows. The electric field at
the $p$-$n$ interface is given by
\begin{equation}
\label{eq:F}
e |F_{pn}| = 2.5\, \hbar v\, \alpha^{1/3}
                 (\rho^\prime_\text{cl})^{2/3}\,,
\end{equation}
where $\rho^\prime_\text{cl} > 0$ is the density gradient at the $p$-$n$
interface computed according to the classical electrostatics.
Equation~(\ref{eq:F}) implies that $e |F_{pn}|$ exceeds a naive estimate
$e |F_{pn}| = \hbar v k_F(\rho_1) /\, D$, where $k_F = \sqrt{\pi
|\rho|}$ is the Fermi wavevector, by a parametrically large factor
$(\alpha k_F D)^{1/3} \gg 1$ (which in practice may approach
$\sim 10$).
The enhancement is caused by the lack of screening at this interface
where the charge density is very low. We demonstrate that
Eq.~(\ref{eq:R_from_F}) is rigorously valid if $\alpha \ll 1$, so that
it is legitimate to substituting $F_{pn}$ from Eq.~(\ref{eq:F}) into
Eq.~(\ref{eq:R_from_F}) to obtain~\cite{Comment_on_ballistic}
\begin{equation}
\label{eq:R}
R = (1.0 \pm 0.1)\, (h / e^2)\, \alpha^{-1/6}\,
    (\rho^\prime_\text{cl})^{-1/3}\,.
\end{equation}
This value of $R$ is parametrically larger than $(\pi / 2) (h / e^2)
\sqrt{k_F(\rho_1) / D}$ that one would obtain from
Eq.~(\ref{eq:R_from_F}) using the aforementioned naive estimate of $e
|F_{pn}|$~\cite{Comment_on_naive}. This, however, only amplifies the
result (considered paradoxical in the early days of quantum
electrodynamics) that larger barriers are more
transparent for Klein tunneling.

Note that Eq.~(\ref{eq:R}) is universal. It is independent of the
number, shape, or size of the external gates that control the profile of
$\rho(x)$ far from the junction. It is instructive however to illustrate
Eq.~(\ref{eq:R}) on some example. Consider therefore a prototypical
geometry depicted in Fig.~\ref{Fig:Gate}. The voltage difference $-V_g$
between graphene and the semi-infinite gate with the edge at $x = x_g$
determines the total density drop $\rho_2 - \rho_1 = -\kappa_0 V_g / 4
\pi e D$. The density $\rho_2$ is assumed to be fixed by other means,
e.g., a global ``backgate'' on the opposite side of the graphene sheet
(not shown in Fig.~\ref{Fig:Gate}). This model is a reasonable
approximation to the available experimental setups~\cite{Huard_07}. An
analytical expression for $\rho_\text{cl}(x)$ follows from the solution
of a textbook electrostatics problem, Eq.~(10.2.51) of
Ref.~\cite{Morse_book}. It predicts that function $\rho_\text{cl}(x)$
crosses zero at the point $x_{pn} = x_g + (D / \pi) [1 + (|\rho_1| /
\rho_2) + \ln(|\rho_1| / \rho_2)]$. Thus, for obvious physical reasons,
the position $x_{pn}$ of the $p$-$n$ interface is gate-voltage
dependent~\cite{Comment_on_shift}. Taking the derivative of
$\rho_\text{cl}$ at $x = x_{pn}$ and substituting the result into
Eq.~(\ref{eq:R}), we obtain
\begin{equation}
  R = \frac{0.7}{\alpha^{1/6}} \frac{h}{e^2}
      \left(1 - \frac{\rho_1}{\rho_2}\right)^{2 / 3}
      \left| \frac{D}{\rho_1} \right|^{1 / 3},
\quad |\rho_1| \gg \frac{1}{D^2}\,.
\label{eq:R_single-gate}
\end{equation}
At fixed $\rho_2$, $R(\rho_1)$ has an asymmetric minimum at $\rho_1 =
-\rho_2$. Away from this minimum, the more dramatic $R(\rho_1)$
dependence (of potential use in device applications) occurs at the
$\rho_1 \to 0$ side where $R$ diverges. The reason for this behavior of
$R$ is vanishing of the density gradient $\rho^\prime_\text{cl}(x)$ at
far left (above the gate). Equation~(\ref{eq:R_single-gate}) becomes
invalid at $|\rho_1| \lesssim 1 / D^2$ where the gradual junction
approximation breaks down. At this point $R \sim (h / e^2) D$.


Let us now turn to the derivation of the general
formula~(\ref{eq:R}). Our starting point is the basic principle of
electrostatics of metals, according to which we can replace the
potential due to the external gates with that created by the fictitious
in-plane charge density $\rho_\text{cl}(x)$. Shifting the origin to $x =
x_{pn}$, we have the expansion $\rho_\text{cl}(x) \simeq
\rho^\prime_\text{cl} x$ for $|x| \ll D$. The {induced} charge density
$\rho(x)$ attempts to screen the external one to preserve charge
neutrality; thus, a $p$-$n$ interface forms at $x = 0$. We wish to
compute the deviation from the perfect screening $\sigma(x) \equiv
\rho_\text{cl}(x) - \rho(x)$ caused by the quantum motion of the Dirac
quasiparticles.


\noindent{\it Thomas-Fermi domain.---} Consider the region $|x|
\gg x_{s}$,
\begin{equation}
\label{eq:x_TF}
x_{s} \equiv (1 / \pi) (\alpha^2 \rho^\prime_\text{cl})^{-1/3}\,.
\end{equation}
At such $x$ the screening is still very effective, $|\sigma(x)| \ll
|\rho_\text{cl}(x)|$ because the local screening length $r_s(x)$ is
smaller than the characteristic scale over which the background charge
density $\rho_\text{cl}(x)$ varies, in this case $\max\{|x|\,, D\}$.
Indeed, the Thomas-Fermi (TF) screening length for graphene
is~\cite{Fogler_xxx} $r_s = (\kappa_0 /\, 2 \pi e^2) (d \mu /\, d \rho)
\sim 1 / \, \alpha \sqrt{|\rho|}$, where $\mu$ is the chemical potential
\begin{equation}
\mu(\rho) = {\rm sign}(\rho)\sqrt{\pi} \hbar v |\rho|^{1/2}
\label{eq:mu}
\end{equation}
appropriate for the 2D Dirac spectrum. Substituting $\rho_\text{cl}(x)$
for $\rho$, we obtain $r_s \sim |\alpha^2 \rho^\prime_\text{cl}
x|^{-1/2}$ at $|x| \ll D$. Therefore, at $|x| \gg x_{s}$ the condition
$r_s \ll |x|$ that ensures the nearly perfect screening is satisfied.

The behavior of the screened potential $V(x)$ and the electric field
$F(x) = -d V /\, d x$ at $|x| \gg x_{s}$ can now be easily calculated
within the TF approximation,
\begin{equation}
\mu[\rho(x)] - e V(x) = 0\,.
\label{eq:TFA}
\end{equation}
It leads to the relation
\begin{equation}
e F(x) \simeq -{\hbar v}
              \sqrt{\pi / 4}\, (\rho^\prime_\text{cl} / |x|)^{1/2}\,,
\quad x_{s} \ll |x| \ll D\,,
\label{eq:F_far}
\end{equation}
which explicitly demonstrates the aforementioned enhancement of $|F(x)|$
near the junction. The TF approximation is valid if
$k_F^{-1}(x) \ll \max\{|x|\,, D\}$. For $\alpha \sim 1$, this criterion
is met if $|x| \gg x_{s}$. For $\alpha \ll 1$, the TF domain
extends down to $|x| = x_\text{TF} \sim \sqrt{\alpha}\, x_{s}$, see
below.


A more formal derivation of the above results is as follows. From 2D
electrostatics~\cite{Morse_book}, we know that
\begin{equation}
\sigma(x) \equiv \rho_\text{cl}(x) - \rho(x)
          = \frac{\kappa_0}{2 \pi^2 e}
            \,\mathcal{P}\!\int
            \frac{d z}{z - x} F(z)\,.
\label{eq:rho_from_F}
\end{equation}
Combined with Eqs.~(\ref{eq:mu})
and (\ref{eq:TFA}), this yields
\begin{equation}
\rho(x) - \rho^\prime_\text{cl} x = 
          \sqrt{\rho^\prime_\text{cl} x_{s}^3}\,\,
          \mathcal{P}\!\int\limits_0^\infty
          \frac{x d z}{z^2 - x^2} \frac{d}{d z} \sqrt{|\rho(z)|}\,.
\label{eq:TFA_iter}
\end{equation}
Here the upper integration limit was extended to infinity, which is
legitimate if $D \gg x_{s}$. The solution for $\rho(x)$ can now be
developed as a series expansion in $1 / x$. The leading correction to
the perfect screening is obtained by substituting $\rho(x) =
\rho^\prime_\text{cl} x$ into the integral, yielding $\sigma(x) /
\rho(x) \simeq (\pi / 4)\, |x_{s} / x|^{3 / 2} $. In accord with the
arguments above, this correction is small at $|x| \gg x_{s}$. Furthermore,
it falls off sufficiently fast with $x$ to ensure that to the order
${\mathcal O}(x_{s} / D)$ the results for $V(x)$ and $\rho(x)$ at the
origin are insensitive to the large-$x$ behavior. In the opposite limit,
$|x| \ll x_{s}$, one can show that
\begin{equation}
\rho_\text{TF}(x) \simeq c^2 \rho^\prime_\text{cl} \frac{x^2}{x_{s}}\,,
\quad
e |F_\text{TF}| \simeq c \pi \hbar v\, \alpha^{1/3}
                 (\rho^\prime_\text{cl})^{2/3}\,,
\label{eq:F_pn_TF}
\end{equation}
where $c \sim 1$ is a numerical coefficient. (The subscripts serve as
a reminder that these results are obtained within the TF approximation.)

\begin{figure}
\centerline{
\includegraphics[width=2.3in]{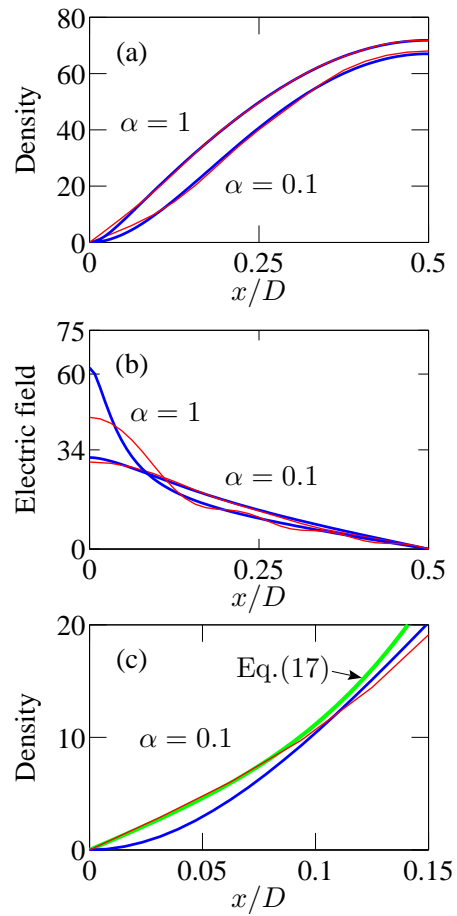}
}
\vspace{0.1in}
\caption{ \label{Fig:results} (Color online) (a) Electron density in
units of $4 / D^2$ for $\alpha = 1$, $\rho_0 = 75$ and $\alpha = 0.1$,
$\rho_0 = 100$. Thicker blue curves are from minimizing the TF
functional, Eqs.~(\ref{eq:E_trial})--(\ref{eq:rho_cl_num}); thinner red
lines are from replacing $E_0$ in this functional by the ground-state
energy of Hamiltonian~(\ref{eq:H}). The $p$-$n$ interface is at $x = 0$.
(b) Magnitude of the electric field in units of $4 \hbar v / e D^2$ for
the same parameters. Numerical values ``34'' and ``60'' are the
predictions of Eq.~(\ref{eq:F}) for the nearby TF (thick blue)
curves~\cite{Comment_on_Friedel}. (c) Enlarged view of the $\alpha =
0.1$ data from the panel (a) and the numerically evaluated
Eq.~(\ref{eq:rho_pn_full}).
}
\end{figure}

Unsuccessful in finding $c$ analytically, we turned to numerical
simulations. To this end we reformulated the problem as the minimization
of the TF energy functional
\begin{gather}
E[V(x)] = E_0 + \int e V(x)
     \left[\frac12 \sigma(x) - \rho_\text{cl}(x)\right] d x\,,
\label{eq:E_trial}\\
E_0 = \frac{e^3}{3 \pi \hbar^2 v^2} \int  |V(x)|^3 d x\,,
\label{eq:E_0}
\end{gather}
where $\sigma(x)$ is defined by Eq.~(\ref{eq:rho_from_F}). The
convolution integral in that equation was implemented by means of a
discrete Fourier transform (FT) over a finite interval $-D \leq x < D$.
Similarly, the integral in Eq.~(\ref{eq:E_trial}) was implemented as a
discrete sum. Since the FT effectively imposes the periodic
boundary conditions on the system, we chose the background charge
density in the form
\begin{equation}
\rho_\text{cl}(x) = \rho_0 \sin(\pi x / D)\,,
\label{eq:rho_cl_num}
\end{equation}
so that the $p$-$n$ interfaces occur at all $x = n D$, where $n$ is an
integer. Starting from the initial guess $\sigma \equiv 0$, the solution
for $\rho(x)$ and $V(x)$ within a unit cell $-D \leq x < D$ was found by
a standard iterative algorithm~\cite{MATLAB}. As shown in
Fig.~\ref{Fig:results}(a), at large $x$ the TF density profile is close
to Eq.~(\ref{eq:rho_cl_num}).
%
%
At small $x$, it is consistent with Eq.~(\ref{eq:F_pn_TF}) using $c =
0.8 \pm 0.05$, cf.~Fig.~\ref{Fig:results}(c).

\noindent{\it Dirac domain.---} Let us now discuss the immediate
vicinity of the $p$-$n$ interface, $|x| < x_\text{TF} \sim \sqrt{\alpha}\, x_{s}$
(the precise definition of $x_\text{TF}$ is given below). At such $x$ the TF
approximation is invalid and instead we have to use the true
quasiparticle wavefunctions to compute $\rho$ and $V$. For a gradual
junction the two inequivalent Dirac points (``valleys'') of
graphene~\cite{Novoselov_04} are decoupled and the wavefunctions can be
chosen to be two-component spinors $\exp(i k_y y)\, [\psi_1(x)\,\,
\psi_2(x)]^T$ (their two elements represent the amplitudes of the
wavefunction on the two sublattices of graphene). Here we already took
advantage of the translational invariance in the $y$-direction and
introduced the conserved momentum $k_y$. The effective Hamiltonian we
need to diagonalize has the Dirac form
\begin{equation}
H = \hbar v (-i \tau_1 \partial_x + \tau_2 k_y) - e V(x)\,,
\label{eq:H}
\end{equation}
where $\tau_1$ and $\tau_2$ are the Pauli matrices. At the end of the
calculation we will need to multiply the results for $\rho(x)$ by the
total spin-valley degeneracy factor $g = 4$.

The solution of this problem can be obtained analytically under the
condition $\alpha \ll 1$. This is possible ultimately because for such
$\alpha$ the electric field is nearly uniform, $|F_{pn} - F(x)| \ll
|F_{pn}|$, inside the strip $|x| \ll x_{s}$. The reason is this strip in
almost empty of charge. Let us elaborate. Since the potential $V(x)$ is
small near the interface and the spectrum is gapless, $\rho(x)$ must be
smooth and have a regular Taylor expansion at $x \to 0$,
\begin{equation}
\rho(x) = a_1 x + a_3 x^3 + \ldots
\label{eq:rho_pn_Dirac}
\end{equation}
Requiring the leading term to match with the TF Eq.~(\ref{eq:F_pn_TF})
at the common boundary $x = x_\text{TF} \sim \sqrt{\alpha}\, x_{s}$ of their
validity, we get $a_1 \sim \sqrt{\alpha}\,\rho^\prime_\text{cl}$. This
means that the net charge per unit length of the interface on the
$n$-side of the junction is somewhat smaller than the TF approximation
predicts, by the amount of $\Delta Q = e \int_0^\infty [\rho(x) -
\rho_\text{TF}(x)] dx \sim \sqrt{\alpha} \rho^\prime_\text{cl} x_\text{TF}^2$.
In turn, the true $|F_{pn}|$ is lower than $|F_\text{TF}|$ by $\sim
\Delta Q /\, \kappa_0 x_\text{TF}$. However for $\alpha \ll 1$ this is only a
small, $\mathcal{O}(\alpha)$ relative correction.

As soon the legitimacy of the linearization $V(x) \simeq -F_{pn} x$ is
established, wavefunctions $\psi_1$ and $\psi_2$ for arbitraty energy
$\epsilon$ are readily found. Since $\epsilon$ enters the Dirac equation
only in the combination $-e V(x) - \epsilon = e F_{pn} (x -
x_\epsilon)$, the energy-$\epsilon$ eigenfunctions are the $\epsilon =
0$ eigenfunctions shifted by $x_\epsilon \equiv \epsilon / (e F_{pn})$
in $x$. In turn, these are known from the literature: they are expressed
in terms of confluent hypergeometric functions $\Phi(a; b;
z)$~\cite{Gradshteyn_Ryzhik}. These solutions were rediscovered multiple
times in the past, both in solid-state and in particle physics. The
earliest instance known to us is Ref.~\cite{Kane_69}; the latest
examples are Refs.~\cite{Cheianov_06} and \cite{Andreev_xxx}. The sought
electron density $\rho(x)$ can now be obtained by a straightforward
summation over the occupied states ($\epsilon \leq 0$), which leads us
to~\cite{Comment_on_Andreev}
\begin{equation}
\rho = \frac{g}{x_\text{TF}^2} \int\! \frac{d k_y}{2 \pi}
\!\!\int\limits_0^x\! \frac{d z}{\pi e^{2 \pi \nu}}
 \left[\left|
       \Phi\left(i \nu; \frac12; -\frac{i z^2}{x_\text{TF}^2}\right)\right|^2
 - \frac12 \right]\,,
\label{eq:rho_pn_full}
\end{equation}
where $\nu = k_y^2 x_\text{TF}^2 / 4$ and $x_\text{TF} \equiv \sqrt{\hbar v /\, |F_{pn}| }
\sim \sqrt{\alpha}\, x_{s}$. This formula is fully consistent with
Eq.~(\ref{eq:rho_pn_Dirac}): the Taylor expansion of the integrand
yields, after a simple algebra, $a_1 = g / (\sqrt{2} \pi^2 x_\text{TF}^3)$, $a_3
= g \sqrt{2} / (3 \pi^3 x_\text{TF}^5)$, {\it etc\/}. Using the known integral
representations of the function $\Phi$~\cite{Gradshteyn_Ryzhik}, one can
also deduce the behavior of $\rho(x)$ at $x \gg x_\text{TF}$. The leading term
is precisely the TF result $\rho_\text{TF}(x) = g x^2 /\, 4 \pi x_\text{TF}^4$.
Therefore, Eq.~(\ref{eq:rho_pn_full}) seamlessly connects to
Eq.~(\ref{eq:F_pn_TF}) at $x \sim x_{s}$. (At such $x$ corrections to
$\rho_\text{TF}(x)$, including Friedel-type
oscillations~\cite{Fogler_unpub}, are suppressed by extra powers of
parameter $\alpha$.) We conclude that for $\alpha \ll 1$ we have
obtained the complete and rigorous solution for $\rho(x)$, $V(x)$, and
$F_{pn}$ [Eq.~(\ref{eq:F})], in particular. As discussed in the
beginning, it immediately justifies the validity of
Eq.~(\ref{eq:R_from_F}) and leads to our result for the ballistic
resistance, Eq.~(\ref{eq:R}). However, in current
experiments $\alpha \sim 1$ and in the remainder of this Letter we
offer a preliminary discussion of what can be expected there.


Since it is the strip $|x| < x_\text{TF}$ that controls the ballistic
transport across the junction~\cite{Cheianov_06}, the constancy of the
electric field in this strip is crucial for the accuracy of
Eq.~(\ref{eq:R_from_F}). This is assured if $\alpha \ll 1$ but at
$\alpha \sim 1$ the buffer zone between $x_\text{TF}$ and $x_{s}$
vanishes, and so we expect $F(x_\text{TF})$ and $F(0) = F_{pn}$
to differ by some numerical factor.

To investigate this question we again turned to numerical simulations.
We implemented a lattice version of the Dirac Hamiltonian by replacing
$-i \partial_x$ in Eq.~(\ref{eq:H}) with a finite difference on a
uniform grid. We also replaced $E_0$ in Eq.~(\ref{eq:E_0}) by the
ground-state energy of $H$, taken with the negative sign: $E_0 =
-L_y^{-1} \sum_{j} {\epsilon_j}/\,{[1 + \exp(\beta \epsilon_j)]}$. Here
$\epsilon_j$ are the eigenvalues of $H$ (computed numerically) and the
$\beta$ is a computational parameter (typically, four orders of
magnitude larger than $1 / \max e |V|$). We have minimized thus modified
functional $E$ by the same algorithm~\cite{MATLAB}, which produced the
results shown in Fig.~\ref{Fig:results}. As one can see, for $\alpha =
0.1$ the agreement between analytical theory and simulations is very
good. However for $\alpha = 1$ we find that $|F_{pn}|$ is approximately
$25\%$ smaller than given by Eq.~(\ref{eq:F}). Note also that for
$\alpha = 1$ the electric field is noticeably nonuniform near the
junction, in agreement with the above
discussion~\cite{Comment_on_Friedel}. Therefore, Eq.~(\ref{eq:R_from_F})
should also acquire some corrections. In principle, we could compute
numerically the transmission coefficients $T(k_y)$ for this more
complicated profile of $F(x)$. However this would not be the ultimate
answer to this problem. Indeed, at $\alpha \sim 1$ electron interactions
are not weak, and so exchange and correlation effects are likely to
produce further corrections to the self-consistent single-particle
scheme we employed thus far, which may be quite nontrivial inside the
Dirac strip $|x| < x_\text{TF}$. We leave this issue
for future investigation.





We are grateful to V.~I.~Falko, D.~S.~Novikov, and B.~I.~Shklovskii for
valuable comments and discussions, to L.~I.~Glazman for a copy of
Ref.~\cite{Kane_69}, to UCSD ACS for support, and to the Aspen Center
for Physics and W.~I.~Fine TPI for hospitality (M.~F.).



\vspace{-0.2in}
\begin{thebibliography}{9}

\bibitem{Novoselov_04}
For a review, see A.~K.~Geim and K.~S.~Novoselov,
Nat.\ Mat.\ {\bf 6}, 183 (2007).

\bibitem{Huard_07} B.~Huard, J.~A.~Sulpizio, N.~Stander, K.~Todd, B.~Yang,
and D.~Goldhaber-Gordon,
Phys.\ Rev.\ Lett.\ {\bf 98}, 236803 (2007);
%
%
B.~\"Ozyilmaz, P.~Jarillo-Herrero, D.~Efetov,
D.~A.~Abanin, L.~S.~Levitov, and P.~Kim,
arXiv:0705.3044;
%
%
J.~R.~Williams, L.~DiCarlo, and C.~M.~Marcus,
arXiv:0704.3487.

\bibitem{Katsnelson_06} M.~I.~Katsnelson, K.~S.~Novoselov, and A.~K.~Geim,
Nat.\ Phys.\ {\bf 2}, 620 (2006).

\bibitem{Cheianov_07} V.~V.~Cheianov, V.~I.~Fal'ko, and B.~L.~Altshuler,
Science\ {\bf 315}, 1252 (2007).

\bibitem{Kane_69} E.~O.~Kane and E.~I.~Blount,
pp. 79--91 in
{\it Tunneling Phenomena in Solids\/}, edited by
E.~Burstein  and S.~Lundqvist (Plenum, New York, 1969).

\bibitem{Cheianov_06} V.~Cheianov and V.~Fal'ko,
Phys.\ Rev.\ B\ {\bf 74}, 041403 (2006).

\bibitem{Comment_on_ballistic} This formula of course neglects
disorder effects that are important in current
experiments~\cite{Huard_07, Fogler_dirt}.

\bibitem{Fogler_dirt} M.~M.~Fogler, L.~I.~Glazman, D.~S.~Novikov, and
B.~I.~Shklovskii, unpublished. 

\bibitem{Comment_on_naive} These incorrect values of $F$ and $R$ could
be inferred from Ref.~\cite{Cheianov_06} if $D$ is incautiously
identified with parameter $d$ therein, as Fig.~1 of that paper
indeed prompts one to do.

\bibitem{Morse_book} P.~M.~Morse and H.~Feshbach,
{\it Methods of Theoretical Physics\/} (McGraw-Hill, New York, 1953).

\bibitem{Comment_on_shift} One can manipulate $x_{pn}$ by the backgate
voltage, which shifts $\rho_\text{cl}(x)$ by a constant
affecting neither $\rho^\prime_\text{cl}(x)$ nor $R$.

\bibitem{Fogler_xxx} See M.~M.~Fogler, D.~S.~Novikov, and B.~I.~Shklovskii,
arXiv:0707.1023
and references therein.

\bibitem{MATLAB} Function \texttt{fminunc} of MATLAB,
\copyright MathWorks, Inc.




\bibitem{Gradshteyn_Ryzhik} I.~S.~Gradshteyn and I.~M.~Ryzhik,
\textit{Table of Integrals, Series, and Products\/}, 6th ed., edited by
A.~Jeffrey and D.~Zwillinger (Academic, San Diego, 2000).

\bibitem{Andreev_xxx} A.~ V.~Andreev,
arXiv:0706.0735.

\bibitem{Comment_on_Andreev} A similar expression was derived in
Ref.~\cite{Andreev_xxx} in the context of carbon nanotube $p$-$n$
junctions. The only difference is that no integration over $k_y$
is present there.

\bibitem{Fogler_unpub} L. M. Zhang and M. M. Fogler, unpublished.

\bibitem{Comment_on_Friedel} The undulations of $F(x)$ seen on the
$\alpha = 1$ curves in Fig.~\ref{Fig:results}(b) may be the
aforementioned Friedel oscillations but we cannot exclude numerical
artifacts either.





\end{thebibliography}
\end{document}